# Current-induced perpendicular effective magnetic field in magnetic heterostructures

Qianbiao Liu,[1] Lijun Zhu[1,2*]

*1. State Key Laboratory for Superlattices and Microstructures, Institute of Semiconductors, Chinese Academy of Sciences, Beijing 100083, China*
*2. College of Materials Science and Opto-Electronic Technology, University of Chinese Academy of Sciences, Beijing 100049, China*
*\*Email: ljzhu@semi.ac.cn*

**Abstract**: Generation of perpendicular effective magnetic field or perpendicular spins ($\sigma_z$) is central for the development of energy-efficient, scalable, and external-magnetic-field-free spintronic memory and computing technologies. Here, we report the first identification and the profound impacts of a significant effective perpendicular magnetic field that can arise from asymmetric current spreading within magnetic microstrips and Hall bars. This effective perpendicular magnetic field can exhibit all the three characteristics that have been widely assumed in the literature to "signify" the presence of a flow of $\sigma_z$, *i.e.*, external-magnetic-field-free current switching of uniform perpendicular magnetization, a sin2$\varphi$-dependent contribution in spin-torque ferromagnetic resonance signal of in-plane magnetization ($\varphi$ is the angle of the external magnetic field with respect to the current, and a $\varphi$-independent but field-dependent contribution in the second harmonic Hall voltage of in-plane magnetization. This finding suggests that it is critical to include current spreading effects in the analyses of various spin polarizations and spin-orbit torques in magnetic heterostructure. Technologically, our results provide perpendicular effective magnetic field induced by asymmetric current spreading as a novel, universally accessible mechanism for efficient, scalable, and external-magnetic-field-free magnetization switching in memory and computing technologies.

## INTRODUCTION

Development of high-density magnetic memory and computing technologies requires energy-efficient and scalable electrical switching of perpendicular magnetization. Microscopically, perpendicular magnetization can be switched by a current of transverse spins ($\sigma_y$) under the assist of an in-plane effective magnetic field along current ($H_x^{\text{eff}}$, to overcome the Dzyaloshinskii–Moriya interaction or to break the switching symmetry)[1], or by an anti-damping spin torque exerted by high-density current of perpendicular spins ($\sigma_z$)[2-10], or by a strong perpendicular effective magnetic field ($H_z^{\text{eff}}$)[11-15]. Since the first method ($\sigma_y$+$H_x^{\text{eff}}$) is hardly energy-efficient or scalable, searching of $\sigma_z$ or $H_z^{\text{eff}}$ in magnetic heterostructures becomes a very hot topic.

Experimentally, it is widely assumed that the presence of $\sigma_z$ could be concluded from a small but sizable sin2$\varphi$ dependent contribution in spin torque ferromagnetic resonance (ST-FMR)[2,7,8,16-18] or a $\varphi$ independent second harmonic Hall voltage response (HHVR)[5,19,20] of an in-plane magnetization ($\varphi$ is the angle of external magnetic field with respect to the current). Presence of $\sigma_z$ is also claimed from the occurrence of external-field-free current switching of a uniform perpendicular magnetization[4,9-15] because the polarization and fieldlike spin-orbit torque (SOT) field of $\sigma_z$, if any, are along the film normal.

In this work, we show that $H_z^{\text{eff}}$ can arise from asymmetric current spreading and thus can widely exist in ST-FMR bars and Hall bars of magnetic heterostructures. While being microscopically distinct from $\sigma_z$, this current-spreading-induced $H_z^{\text{eff}}$ shows all the three characteristics that were widely assumed in the literature to signify the presence of $\sigma_z$, *i.e.*, it can enable external-field-free current switching of perpendicular magnetization and contributes to HHVR and ST-FMR signals of in-plane magnetization in analogue to the fieldlike SOT of $\sigma_z$. Neglect of asymmetric current spreading can lead to erroneous analyses of various spin polarizations and SOTs.

**Sample Characterizations**

For this work, a series of Pt 4/Py 3.3-9.4, Pt 4/FeCoB 2.8-9.6, Pt 4/Ni 2.4-9.2, Pt 4/Co 1.7, and Pt 4/ FeTb 7.5 bilayers are sputter-deposited on oxidized Si substrates (the numbers are layer thicknesses in nanometer, FeCoB = $Fe_{60}Co_{20}B_{20}$, Py = $Ni_{81}Fe_{19}$, FeTb = $Fe_{65}Tb_{35}$). Each sample is protected by a MgO 1.6 /Ta 1.6 bilayer that is fully oxidized upon exposure to the atmosphere[21]. The samples are patterned into microstrips and Hall bars by photolithography and ion milling, followed by deposition of Ti 5/Pt 150 as the contacts for ST-FMR, HHVR, and switching measurements.

**Effective Perpendicular Magnetic Field within Magnetic Strips**

We first perform ST-FMR measurements on the Pt 4/Py 9.4 microstrips using the nominally symmetric 3-terminal contact configuration in Fig. 1(a). The symmetric (*S*) and anti-symmetric (*A*) components of the ST-FMR responses for the Pt 4/Py 9.4 are plotted in Figs. 1(b) and 1(c) as a function of *φ* (see Ref. 28 and Supplementary materials for the method how to determine the *S* and *A*). Considering a magnetic strip interacting with a spin current with arbitrary spin polarization $\boldsymbol{\sigma} = (\sigma_x, \sigma_y, \sigma_z)$, the *S* and *A* values should vary with *φ* following[2,22,23]:

$$S = S_{\text{DL},y} \sin2\varphi\cos\varphi + S_{\text{DL},x} \sin2\varphi\sin\varphi + S_{\text{FL},z} \sin2\varphi + S_{\text{SP}}\sin\varphi, \quad (1)$$

$$A = A_{\text{FL},y} \sin2\varphi \cos\varphi + A_{\text{FL},x} \sin2\varphi \sin\varphi + A_{\text{DL},z} \sin2\varphi. \quad (2)$$

The four terms of Eq. (1) are contributions of the dampinglike SOT of $\sigma_y$, the dampinglike SOT of $\sigma_x$, $H_z^{\text{eff}}$ (fieldlike SOT of $\sigma_z$ and others), and spin pumping, respectively. Equation (2) includes the contributions from the sum of the fieldlike SOT of $\sigma_y$ and the transverse Oersted field, the fieldlike SOT of $\sigma_x$, and the dampinglike SOT of $\sigma_z$. The fits of the *S* and *A* data to Eqs. (1) and (2) untangle different contributions and yield the values of $S_{\text{DL},y}$, $S_{\text{DL},x}$, $S_{\text{FL},z}$, $S_{\text{SP}}$, $A_{\text{FL},y}$, $A_{\text{FL},x}$, and $A_{\text{DL},z}$. As expected, there is a predominant contribution from $\sigma_y$ ($S_{\text{DL},y}$, $A_{\text{FL},y}$) but no indication of $\sigma_x$ ($S_{\text{DL},x} = 0$, $A_{\text{FL},x} = 0$, see Figs. 1(b) and 1(c)).

However, it is striking that such heavy metal/ferromagnet (HM/FM) samples exhibit a non-negligible $S_{\text{FL},z}$ term and thus a non-zero $H_z^{\text{eff}}$, in the case of the 3-terminal geometry (Fig. 1(a)). This $H_z^{\text{eff}}$ is unlikely to arise from $\sigma_z$ because we only measure a negligible dampinglike SOT of $\sigma_z$ (i.e., $A_{\text{DL},z} \approx 0$). This is reasonable because the sputter-deposited polycrystalline bilayers do not have a long-range lateral crystal or magnetic symmetry breaking.

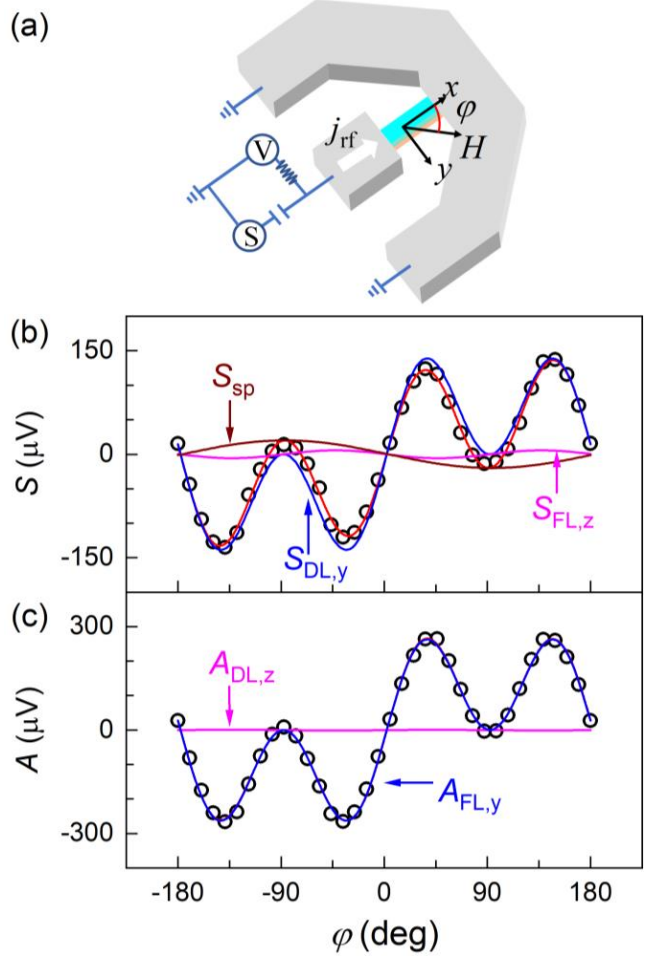

**Fig. 1.** ST-FMR results of the Pt 4/Py 9.4 strip measured using the nominally symmetric 3-terminal contact configuration ($W$=10 μm, $L$=20 μm, the rf power is 15 dBm). (a) Schematic of the 3-terminal ST-FMR measurement configuration. $\varphi$ dependence of (b) $S$ and (c) $A$ from the 3-terminal ST-FMR measurements. The solid curves in (b) represent the $S_{FL,z}$, $S_{DL,y}$, and $S_{sp}$ components as determined from fit of the data to Eq.(1). The solid curves in (c) represent the $A_{DL,z}$ and $A_{FL,y}$ components as determined from fit of the data to Eq.(2).

We also find no obvious correlation between $S_{FL,z}$ and the interfacial SOC strength for the Pt/FM samples. First, we find a strong $S_{FL,z}$ component in control samples Cu 2/Py 9.4/Cu 2 and Cu 2/FeCoB 9.6/Cu 2 with negligible SOC (Figs. 2(a) and 2(b)). As we show in Fig. 2(c), the $S_{FL,z}/S_{DL,y}$ ratios for the Pt/Py, Pt/FeCoB, and Pt/Ni samples, with similar layer thicknesses, do not scale with the interfacial perpendicular magnetic anisotropy energy density ($K_s$) of the Pt/FM interfaces (as determined in Supplementary Materials), with $K_s$ reflecting the interfacial SOC strength[24,25]. These observations indicate that the $H_z^{eff}$ here is not induced by any SOC-related precession/rotation scattering[4,5]. The $S_{FL,z}/A_{FL,y}$ ratio of a given device is found to be largely independent of the power of the rf current (Fig. 2(d)), excluding any relevance of the $H_z^{eff}$ to thermal effects. Note that the "hidden" vertical symmetry breaking that leads to bulk SOT of $\sigma_y$ in ferromagnets[21,26] and ferrimagnets[27] cannot explain the $H_z^{eff}$ because the latter requires a lateral rather than vertical asymmetry.

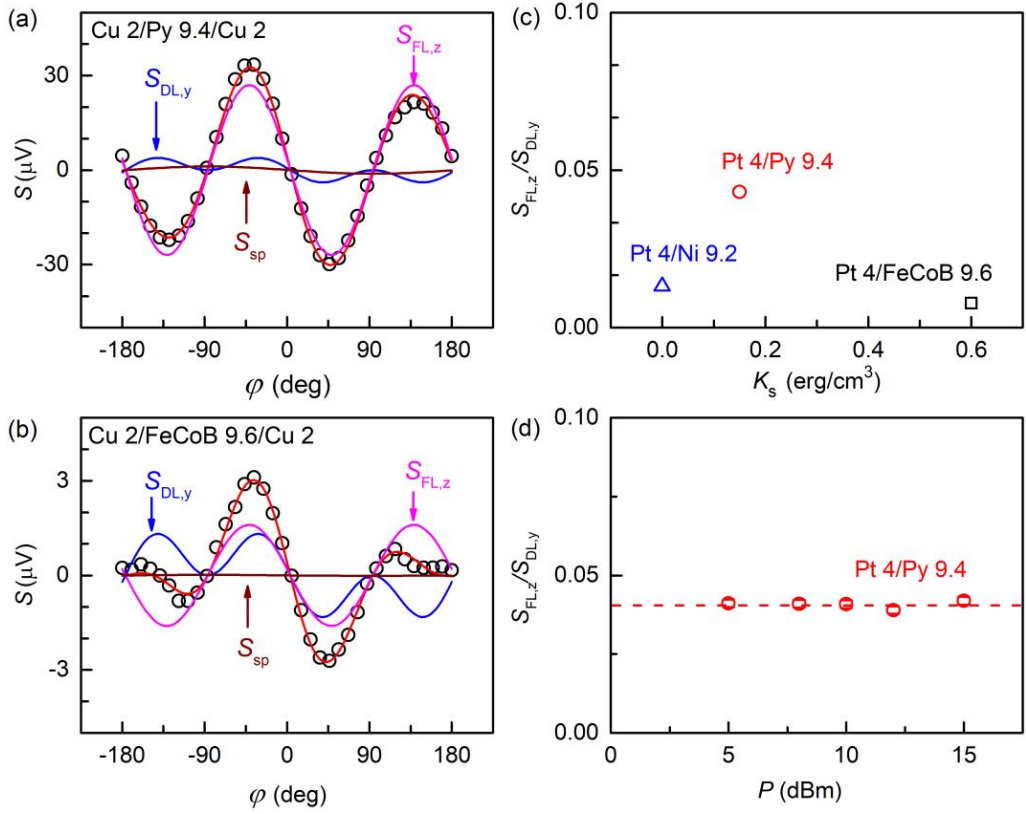

Fig. 2. $\varphi$ dependence of $S$ (a) for the Cu 2/Py 9.4/Cu 2 strip and (b) for the Cu 2/FeCoB 9.6/Cu 2 strip ($W$=10 μm, $L$=20 μm). The solid curves represent the $S_{FL,z}$, $S_{DL,y}$, and $S_{sp}$ components as determined from fit of the data to Eq.(1). (c) $S_{FL,z}/S_{DL,y}$ for the Pt 4/Py 9.4, Pt 4/FeCoB 9.6, and Pt 4/Ni 9.2 plotted as a function of interfacial perpendicular magnetic anisotropy energy density ($K_s$) of the Pt/FM interfaces. (d) Dependence on the rf power of $S_{FL,z}/S_{DL,y}$ for the Pt 4/Py 9.4. The data are from 3-terminal ST-FMR measurements. In (a)-(c), the rf power is 15 dBm.

Next, we show that the $H_z^{\text{eff}}$ is microscopically a perpendicular Oersted field arising from asymmetric current spreading near the contacts of ST-FMR devices. First, $S_{FL,z}$ for the same Pt/Py device becomes a factor of 6 greater when the contact is switched from 3-terminal ST-FMR configuration (nominally symmetric, Fig. 1(a)) to strongly asymmetric 2-terminal ones (Figs. 3(a) and 3(d)). $S_{FL,z}$ also reverses sign when contact geometry is switched from that in Fig. 3(a) to the one in Fig. 3(d), which is in contrast to $S_{DL,y}$ that remains essentially the same in both sign and magnitude.

One of the features of current spreading near the contacts should be an apparent dependence of the resultant $H_z^{\text{eff}}$ on the width $W$ and length $L$ of the ST-FMR device, because the lateral current density asymmetry should decrease (increase) with reducing $W$ ($L$). This can be quantitatively checked by using the ratio of $S_{FL,z}/S_{DL,y}$ to represent the relative strength of $H_z^{\text{eff}}/H_{DL,y}$. Indeed, the measured value of $S_{FL,z}/S_{DL,y}$ for the Pt 4/Py 9.4 devices decreases substantially as $W$ decreases ($L$= 20 μm, Fig. 4(a)) and increases as $L$ decreases ($W$ = 10 μm, Fig. 4(a)). As shown in Fig. 4(b), the $S_{FL,z}/S_{DL,y}$ ratio increases with the FM thickness approximately following a parabolic function, which can be understood by in-plane current spreading near the contacts (Supplementary Materials).

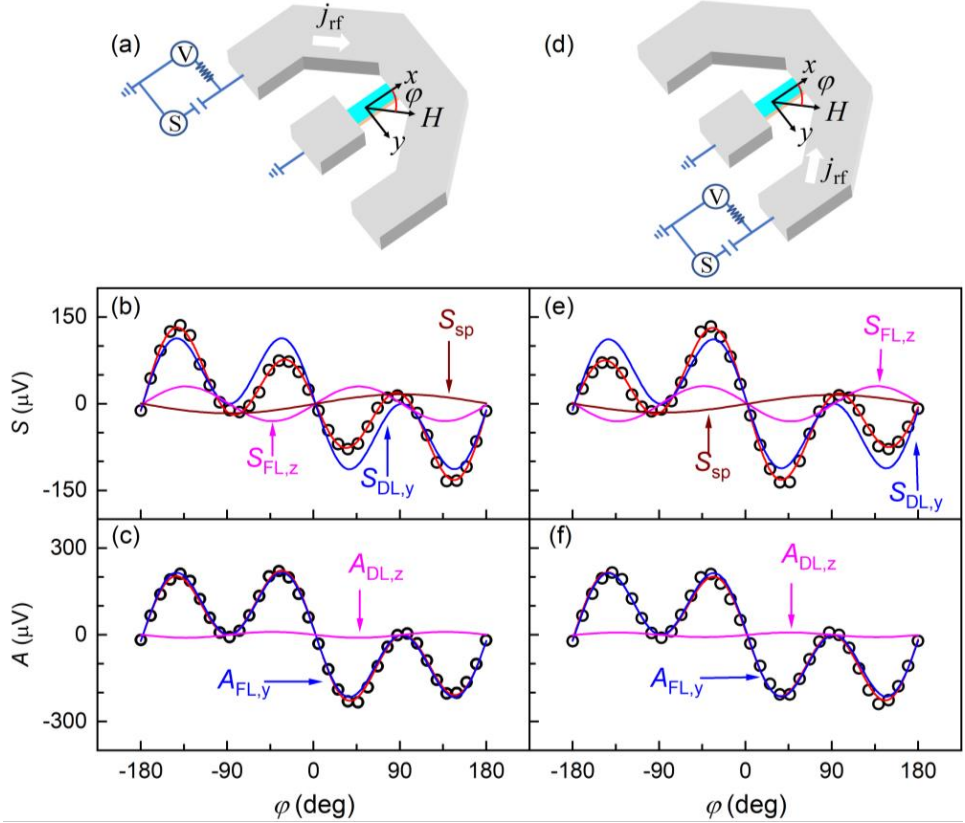

**Fig. 3.** ST-FMR results of the Pt 4/Py 9.4 strip measured using asymmetric 2-terminal configurations ($W$=10 μm, $L$=20 μm, the rf power is 15 dBm). Schematics showing the 2-terminal ST-FMR configurations with rf current being injected from (a) left and (d) right contact arms, respectively. $\varphi$ dependence of (b) $S$ and (c) $A$ from 2-terminal ST-FMR measurements with contact configuration in (a). $\varphi$ dependence of (e) $S$ and (f) $A$ from 2-terminal ST-FMR measurements with contact configuration in (d). The solid curves in (b) and (e) represent the $S_{FL,z}$, $S_{DL,y}$, and $S_{sp}$ components as determined from fit of the data to Eq.(1). The solid curves in (c) and (f) represent the $A_{DL,z}$ and $A_{FL,y}$ components as determined from fit of the data to Eq.(2).

**Effects on spin-torque ferromagnetic resonance analysis**

Now we discuss how the $H_z^{\text{eff}}$ affects the estimation of the efficiency of the dampinglike SOT of $\sigma_y$. Taking into account the asymmetric current spreading effect, we define a FMR efficiency using the values of $S_{DL,y}$ and $A_{FL,y}$ from angle-dependent FMR measurements (Eq. (1) and Eq. (2)) as

$$\xi_{\text{FMR},y} \equiv \frac{S_{DL,y}}{A_{FL,y}} \frac{e\mu_0 M_s t_{FM} t_{HM}}{\hbar} \sqrt{1 + \frac{4\pi M_{\text{eff}}}{H_r}}, \quad (3)$$

$\xi_{DL,y}^{j}$ would be the inverse intercept in the linear fit of $1/\xi_{\text{FMR},y}$ vs $1/t_{FM}$ [28,29]. Here, $e$ is the electron charge, $\hbar$ the reduced Planck constant, $\mu_0$ the vacuum permeability, $t_{HM}$ the thickness of the HM, and $4\pi M_{\text{eff}}$ the effective demagnetization field of the FM estimated from Kittel's equation (Supplementary Materials). The saturation magnetization ($M_s$) is determined by vibrating sample magnetometry to be $370 \pm 30$ emu/cm$^3$ for the Ni, $705 \pm 50$ emu/cm$^3$ for the Py, and $1180 \pm 40$ emu/cm$^3$ for the FeCoB.

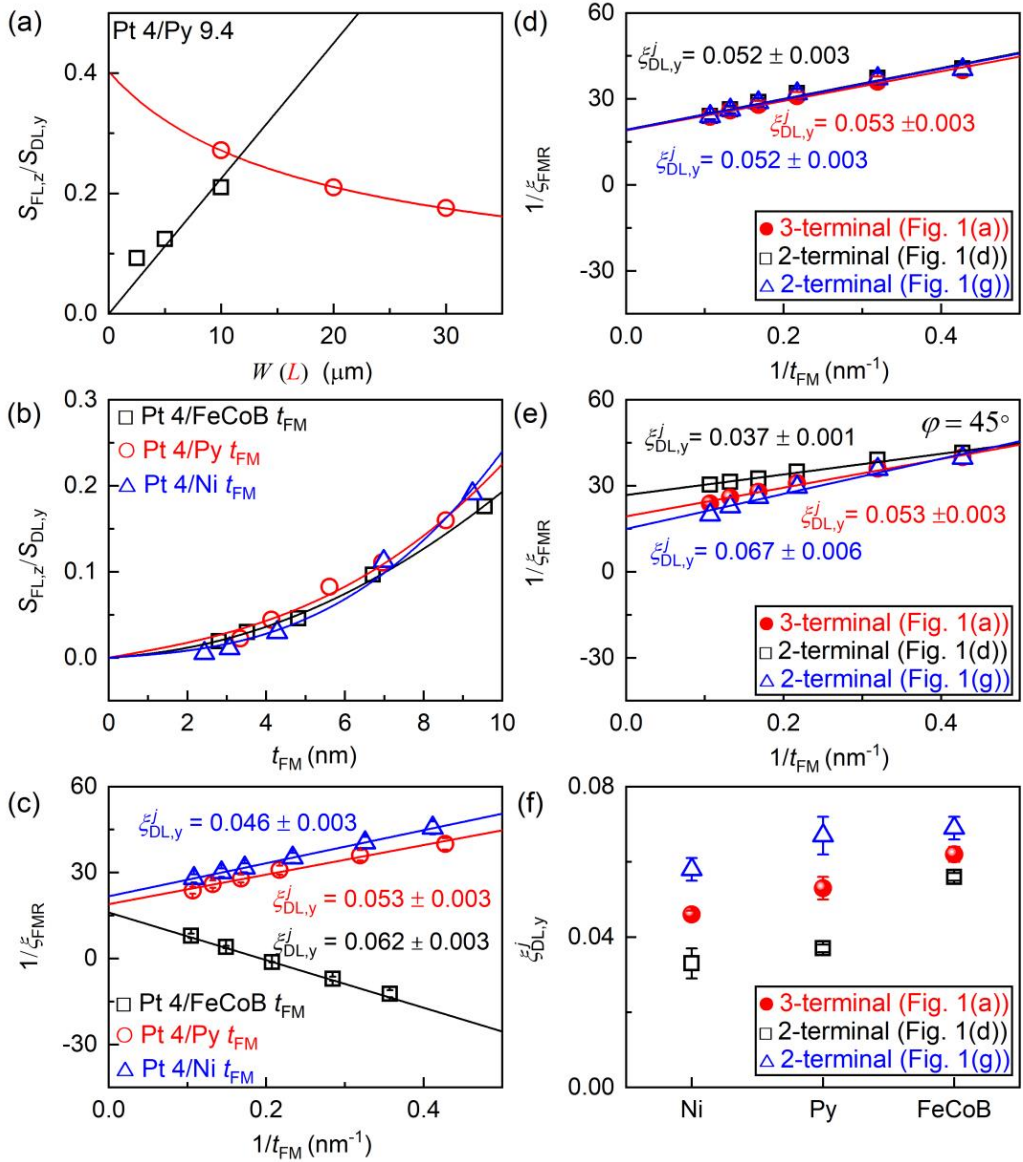

**Fig. 4** (a) $S_{FL,z}/S_{DL,y}$ for the Pt 4/Py 9.4 strips with different width $W$ ($L$=20 μm) and length $L$ ($W$=10 μm) as measured in the 3-terminal ST-FMR configuration in Fig. 1(b). (b) The FM thickness dependence of $S_{FL,z}/S_{DL,y}$ for Pt 4/Py $t_{Py}$, Pt 4/Ni $t_{Ni}$, and Pt 4/FeCoB $t_{FeCoB}$ devices ($W$=10 μm, $L$=20 μm) as measured using the 2-terminal ST-FMR configuration in Fig. 1(b). (c) $1/\xi_{FMR}$ vs $1/t_{FM}$ for Pt 4/Py $t_{Py}$, Pt 4/Ni $t_{Ni}$ and Pt 4/FeCoB $t_{FeCoB}$. $1/\xi_{FMR}$ vs $1/t_{FM}$ for Pt 4/Py $t_{Py}$ as measured using ST-FMR measurements with three different contact configurations: (d) angle dependent analysis (-180° ≤ $\varphi$ ≤ 180°) and (e) fixed-angle analysis at $\varphi$ = 45°. (f) Over- and under-estimation of $\xi_{DL,y}^{j}$ in 2-terminal ST-FMR measurements at $\varphi$ = 45° compared to 3-terminal measurements.

Figure 4(c) shows the $1/\xi_{FMR,y}$ vs $1/t_{FM}$ results from 3-terminal ST-FMR measurements, from which $\xi_{DL,y}^{j}$ is determined to be 0.062 ± 0.003 for Pt/FeCoB, 0.053 ± 0.003 for Pt/Py, and 0.046 ± 0.003 for Pt/Ni. The slight variation of $\xi_{DL,y}^{j}$ of these samples is attributed to the change of resistivity ($\rho_{xx}$) and thus the spin Hall ratio ($\theta_{SH}$= $\sigma_{SH}\rho_{xx}$, where $\sigma_{SH}$ is the spin Hall conductivity of the transverse spins) of the Pt layers [30]. As we determine by measuring the conductance enhancement of the corresponding stacks ($t_{Pt}$ = 4 nm) relative to the control $t_{Pt}$ = 0 nm stack, $\rho_{xx}$ of the 4 nm Pt layer is 42 μΩ cm, 32 μΩ cm, and 29 μΩ cm in contact with FeCoB, Py, and Ni, respectively. Note that the spin transparency of the Pt/FM interfaces should be close in these samples because the spin-mixing conductance of Pt/FM interfaces is insensitive to the type of the FM [25] and there is negligible spin memory loss[31] as indicated by the small interfacial magnetic anisotropy energy density (Fig. 2(c)). Importantly, we find that, regardless of the contact configuration, $\xi_{DL,y}^j$ can be determined from such angle- and thickness-dependent ST-FMR measurements (Fig. 4(d)).

However, as we show in Figs. 4(e) and 4(f), a 2-terminal ST-FMR measurement at a single angle (e.g., $\varphi$ = 45°) leads to a substantial under- or over-estimation of $\xi_{DL,y}^j$, depending on the current direction. This observation is consistent with a previous study that the $S_{DL,y}/A_{FL,y}$ ratio in 2-terminal ST-FMR measurement deviates from that from $\varphi$-dependent 3-terminal measurements[32]. This is likely a universal issue for 2-terminal ST-FMR measurements in which a significant $H_z^{eff}$ exists due to asymmetric current spreading[33].

**Effects on second harmonic Hall voltage of in-plane magnetization**

We show below that such asymmetric current spreading also induces $H_z^{eff}$ in Hall bar devices (Figs. 5(a)-5(f)). Within the approximation of an in-plane magnetized macrospin, the angle dependence of the second HHVR ($V_{2\omega}$) under a sinusoidal electric field $E$ = 25.4 kV/m and an in-plane magnetic bias field ($H$) follows[19,20,34,35]:

$$V_{2\omega} = V_{DL,y+ANE} \cos\varphi + V_{FL,y+Oe} \cos\varphi\cos2\varphi + V_{DL,z} \cos2\varphi + V_{FL,z+Oe} + V_{PNE}\sin2\varphi \quad (4)$$

with $V_{DL,y+ANE} = V_{AHE}H_{DL,y}/2(H_k+H) + V_{ANE}$, (5)

$V_{FL,y+Oe} = -V_{PHE}H_y^{eff}/2H$, (6)

$V_{DL,z} = -V_{PHE}H_{DL,z}/2H$, (7)

$V_{FL,z+Oe} = V_{AHE}H_z^{eff}/2(H_k+H)$, (8)

Here, $V_{PHE}$, $V_{AHE}$, $V_{PNE}$, and $V_{ANE}$ are the planar Hall voltage, the anomalous Hall voltage, the planar Nernst voltage, and the anomalous Nernst voltage; $V_{DL,y+ANE}$ is the contributions of dampinglike SOT field of $\sigma_y$ ($H_{DL,y}$) and $V_{ANE}$; $V_{FL,y+Oe}$ is the contributions of fieldlike SOT field of $\sigma_y$ and the transverse Oersted field; $V_{DL,z}$ is contributions of the dampinglike SOT field of $\sigma_z$ ($H_{DL,z}$), and $V_{FL,z+Oe}$ is contributions of the fieldlike SOT field of $\sigma_z$ ($H_{FL,z}$) and the asymmetric spreading current induced $H_z^{eff}$. Fits of the $V_{2\omega}$ data to Eq. (4) yield the values of $V_{DL,y+ANE}$, $V_{FL,y+Oe}$, $V_{DL,z}$, and $V_{FL,z+Oe}$ for the Pt 4/Co 1.7 at each magnitude of $H$ (Figs. 5(b), and 5(e)). According to Eqs. (5)-(8), the slopes of the linear fits of $V_{DL,y+ANE}$ vs $V_{AHE}/2(H_k+H)$, $V_{FL,y+Oe}$ vs $V_{PHE}/2H$, $V_{DL,z}$ vs $V_{PHE}/2H$, $V_{FL,z+Oe}$ vs $V_{AHE}/2(H_k+H)$ give the values of $H_{DL,y}$, $H_y^{eff}$, $H_{DL,z}$, and $H_z^{eff}$. Here, the anisotropy field of the in-plane magnetization, $H_k$, is estimated using the out-of-plane saturation field.

As we show in Figs. 5(c) and 5(f), the SOT fields for $\sigma_y$ and $\sigma_z$ are similar for the two contact configurations ($H_{DL,y} \approx$ 15 Oe, $H_{FL,y} \approx$ -1.1 Oe, $H_{DL,z} \approx$ 0), while $H_z^{eff}$ is small (-1.7 $\pm$ 1.3 Oe) in the nominally symmetric contact configuration and much larger (-6.0 $\pm$ 0.9 Oe) in the more asymmetric configuration. The high sensitivity of $H_z^{eff}$ to the contact configuration also unambiguously indicates that the $H_z^{eff}$ in the Hall bars is an extrinsic current spreading effect rather than the fieldlike torque of $\sigma_z$. We expect that, similar to the ST-FMR case (Fig.

4(a)), such asymmetric current spreading near the electrical contacts gets more significant for Hall bars that are wider and/or shorter than the ones in this study. As a result, angle-dependent HHVR measurements are required to reliably determine the SOT efficiencies. For instance, the measurement that records $V_{2\omega}$ as a function of a swept in-plane field $H$ along charge current direction ($\varphi = 0°$) would result in erroneous value of $H_{DL,y}$ because, as Eqs.(5) and (8) suggest, the second HHVR of $H_z^{\mathrm{eff}}$ and the dampinglike SOT of $\sigma_y$ have the same form in the field dependence. The *symmetric* current shunting into the voltage detection arm pairs of Hall bars[36] is not involved with the presence and the analysis of $H_z^{\mathrm{eff}}$.

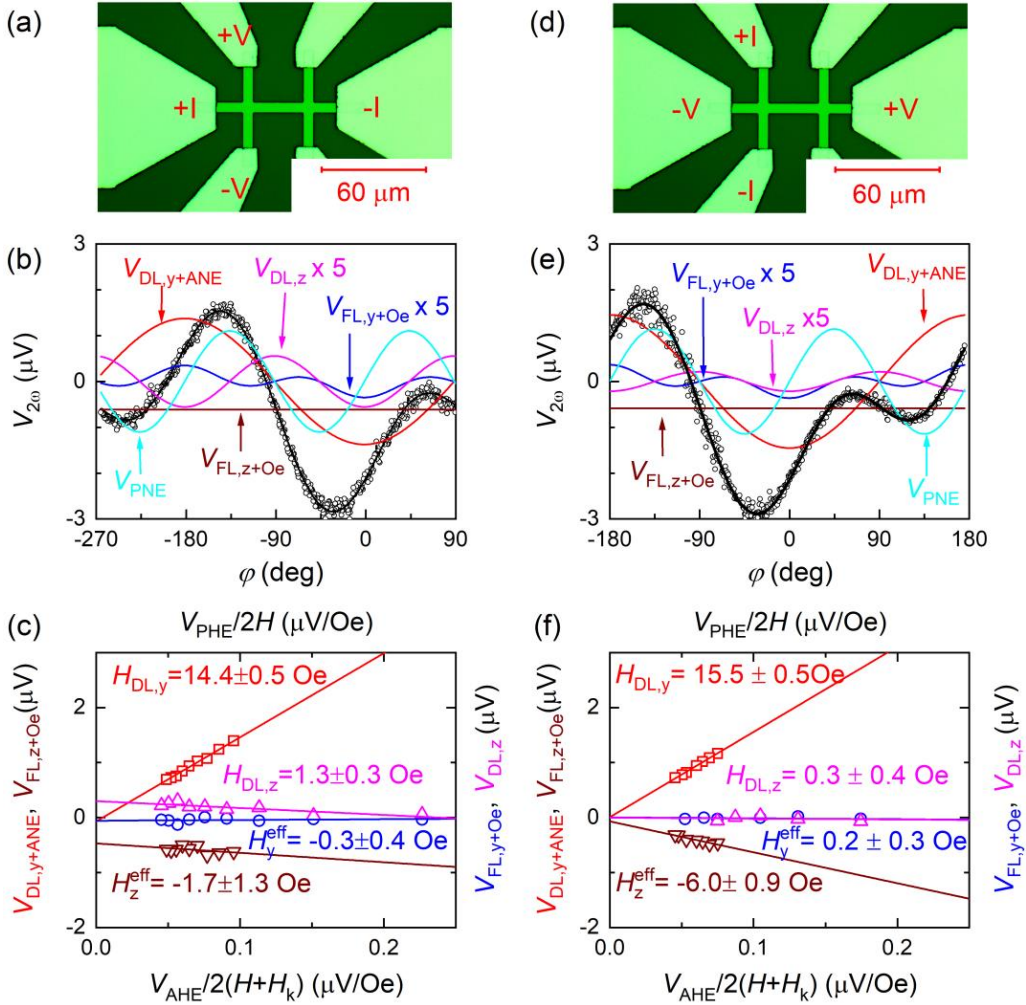

**Fig. 5** (a) HHVR measurement configuration with current injected into a 5×60 µm² strip from nominally symmetric contacts. (b) $\varphi$ dependence of $V_{2\omega}$ for the Pt 4/Co 1.7 ($H$ = 1000 Oe). (c) $V_{DL,y+ANE}$ vs $V_{AHE}/2(H+H_k)$, $V_{FL,z+Oe}$ vs $V_{AHE}/2(H+H_k)$, $V_{FL,y+Oe}$ vs $V_{PHE}/2H$, $V_{DL,z}$ vs $V_{PHE}/2H$ measured with contact configuration in (a). (d) HHVR measurement configuration with current injected from more asymmetric contact into a 5×40 µm² strip. (e) $\varphi$ dependence of $V_{2\omega}$ for the Pt 4/Co 1.7 ($H$ = 1000 Oe). (f) $V_{DL,y+ANE}$ vs $V_{AHE}/2(H+H_k)$, $V_{FL,z+Oe}$ vs $V_{AHE}/2(H+H_k)$, $V_{FL,y+Oe}$ vs $V_{PHE}/2H$, and $V_{DL,z}$ vs $V_{PHE}/2H$ measured with contact configuration in (d). In (b) and (e), the solid curves refer to different contributions as determined from the fits of data to Eq. (4). For clarity, the values of $V_{FL,y+Oe}$ and $V_{DL,z}$ are multiplied by 5 in (b) and (e).

**External-magnetic-field-free switching of uniform perpendicular magnetization**

Finally, we demonstrate that the effective perpendicular magnetic field induced by asymmetric current spreading can enable external-magnetic-field-free current switching of perpendicular magnetization. As expected, in-plane current cannot switch a magnetic bilayer of Pt 5/FeTb 7.5 bilayer ($M_s \approx 350$ emu/cm$^3$, $H_k \approx 30$ kOe, coercivity $H_c \approx 0.5$ kOe) in absence of an assisting external magnetic field, in the configuration of Fig. 6(a) in which the asymmetric current spreading is weak (the current channel is 60 μm long and the current leads have symmetric shapes). When the contact configuration is changed to a more asymmetric one, e.g., the one in Figs. 6(b) and 6(c) (the current channel is 40 um long in Fig. 6(b) and 65 um long in Fig. 6(c) and the current leads have asymmetric shapes), in-plane current can switch the same Pt 5/FeTb 7.5 bilayer at current density of $\approx 1.8 \times 10^7$ A/cm$^2$ without need of an external magnetic field. Here, since $H_z^{\text{eff}}$ is much smaller than the coercivity of the FeTb layer, the observed magnetization switching is driven mainly by the transverse spins generated by the spin Hall effect of Pt and assisted by $H_z^{\text{eff}}$. While the deterministic magnetization switching of the Pt 5/FeTb 7.5 sample is not full yet (volume percentage of 20% or 30%) because some of the magnetic domains appear to be pinned more firmly, a full switching is possible for samples with enhanced asymmetric current spreading and/or with reduced pinning field.

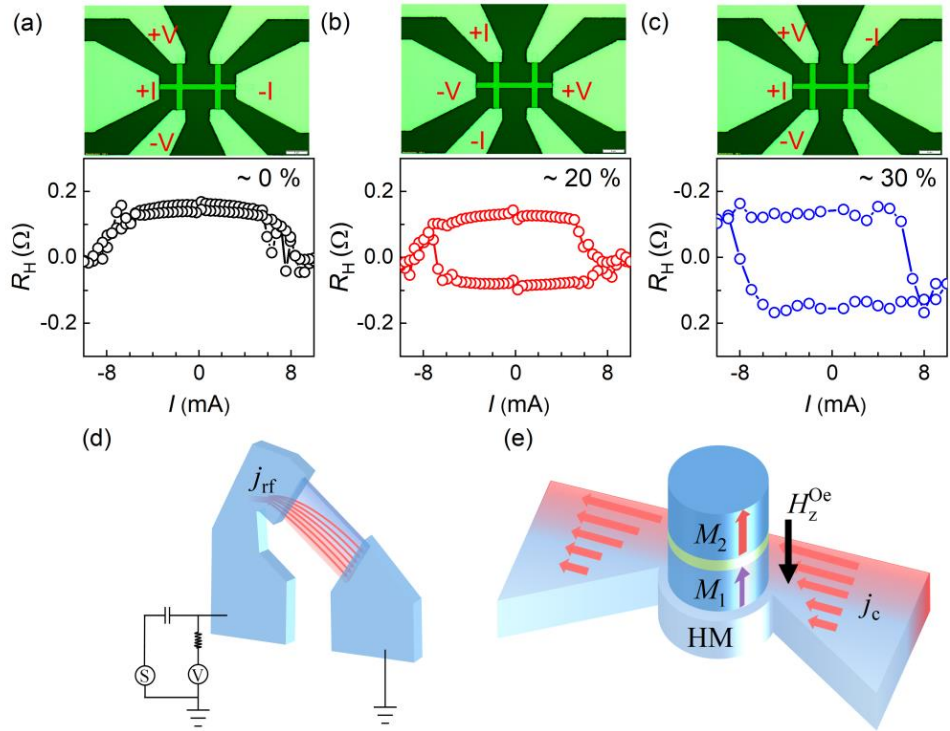

**Fig. 6** Magnetization switching. (a-c) Three different measurement configurations and the Hall resistance for Pt 5/FeTb 7.5 bilayers ($M_s \approx 350$ emu/cm$^3$, $H_k \approx 30$ kOe, $H_c \approx 0.5$ kOe) plotted as a function of bias current in the Pt layer, suggesting occurrence of deterministic magnetization switching in asymmetric configurations ((b) and (c)) but not in the symmetric configurations ((a)). The numbers 0%, 20%, and 30% denote the estimated volume percentages of the switched magnetic domains. (d) Schematic depict of strongly asymmetric current spreading in a "2-terminal" ST-FMR device. (e) Schematic of a SOT-MRAM device with asymmetric current channel.

## CONCLUSION

We have shown that a perpendicular effective magnetic field can widely exist in magnetic micron devices due to asymmetric current spreading and vary with the geometries of devices and measurements. This perpendicular effective field exhibits the characteristics that were widely assumed in the literature to "signify" the presence of $\sigma_z$, *i.e.*, it can enable external-magnetic-field-free current switching of perpendicular magnetization, and make a $\sin2\varphi$ dependent contribution in the ST-FMR signal of in-plane magnetization and a $\varphi$-independent but field-dependent contribution in the second HHVR of in-plane magnetization. This implies that these characteristics are not reliable indication for perpendicular spins ($\sigma_z$) and that determination of SOTs of transverse spins ($\sigma_y$) from ST-FMR and HHVR analyses requires careful angle-dependent measurements, especially those on low-length/width-ratio magnetic strips and/or with asymmetric electrical contacts (e.g. 2-terminal ST-FMR devices similar to that in Fig. 6(d)). Neglect of a significant current spreading causes incorrect conclusions about the presence of $\sigma_z$ and the strength and anisotropy of SOTs of $\sigma_y$ from ST-FMR, HHVR, and magnetization switching analyses. This work also provides a unified solution to reliably measure SOTs of spin polarizations, regardless of the contact configurations, from careful angle dependent ST-FMR and HHVR measurements that have taken in account current spreading. Distinct from thickness-wedged samples[37] and low-symmetry $NbSe_2$ single crystal with anisotropic resistivity[22], such perpendicular effective magnetic field induced by asymmetric current spreading within uniform magnetic heterostructures provides a new, universally accessible mechanism for efficient, scalable and external-field-free switching of perpendicular magnetization in nonvolatile memory and computing technologies (see Fig. 6(e) for a possible integration scheme of a SOT-MRAM device with optimized contact geometry to enhance the spreading Oersted field).

## SUPPLEMENTARY MATERIALS
See the supplementary materials for more details on the determination of the *S* and *A* components of spin-torque ferromagnetic resonance spectrum, interfacial perpendicular magnetic anisotropy energy density, and more discussions on in-plane current spreading near the contacts.

## ACKNOWLEDGEMENTS
This work is supported by the Strategic Priority Research Program of the Chinese Academy of Sciences (XDB44000000).

## AUTHOR DECLARATIONS
**Conflict of Interest:** The authors have no conflicts to disclose.

## DATA AVAILABILITY
The data that support this study are available from the corresponding authors upon reasonable request.